# Weight a Minute: Understanding Variability in PATE Estimates Across Target Populations

Authors: William Stewart[1], Carly L. Brantner[1], Elizabeth A. Stuart[3], Laine Thomas[1]

[1]Department of Biostatistics & Bioinformatics, Duke University School of Medicine, Durham, NC, USA

[2]Duke Clinical Research Institute, Durham, NC, USA

[3]Department of Biostatistics, Johns Hopkins Bloomberg School of Public Health, Baltimore, MD, USA

## Abstract

### Background

Despite techniques to optimize recruitment and reach individuals representing various geographic areas and phenotypes, clinical study populations often differ meaningfully from broader populations to which results are intended to generalize. Weighting methods such as inverse probability of sampling weights (IPSW) are commonly implemented to improve generalizability by reweighting study participants to resemble a chosen target population on observed covariates, thereby estimating the population average treatment effect (PATE). Yet the performance of this approach is driven less by the estimator itself, and more by the data data representing the target population, which determines both the estimand and how well it can be recovered. In this sense, limitations arise from the population definition and available data rather than the weighting method. Despite this, there is limited guidance on how the choice of target population affects IPSW estimator performance in practice.

### Methods

To provide such guidance and demonstrate the impact of target population selection on the accuracy of IPSW estimators of the PATE, we conducted a simulation study grounded in empirical covariate distributions from real-world data sources, including the US Census,

PCORnet, Diabetes Collaborative Registry, and real studies on diabetic populations. These data sources span a continuum of representativeness from highly selective analytic samples to broadly inclusive populations and thus represent a spectrum of possible target populations that researchers may use for generalizing study estimates. We quantified the bias of IPSW PATE estimates in each candidate target population relative to a defined reference population, comparing weighting models that included complete and partial covariate sets across scenarios with a varying number of treatment effect modifiers.

**Results**

Our results demonstrate that the bias we observe is not driven by IPSW itself, but by misalignment between the policy-relevant target population and the datasets available to operationalize that population in practice. Bias magnitude increased with greater imbalance between the true target population and the candidate target population as measured by standardized mean differences (SMD).

**Conclusions**

These findings highlight that weighting methods cannot compensate for a poorly representative target population and therefore the dangers in naively applying IPSW. Valid generalization depends on selecting a target population that is the most robust representation of the intended population for inference, in addition to applying appropriate statistical adjustments.

**Background**

Clinical studies are designed to estimate causal effects under controlled or well-defined conditions. Yet, study populations often differ meaningfully from broader populations to which results are intended to generalize. Differences in relevant covariates that modify treatment effect create meaningful gaps between analytic study samples and the target populations for which evidence is needed (1-3). When these gaps exist, important subgroups may be underrepresented in clinical studies, leading to treatment effects that do not generalize and potentially undermining public trust in clinical evidence (1-3).

To bridge the gap between analytic and target populations, analysts frequently apply statistical generalizability methods such as outcome modeling, and weighting-based approaches including inverse probability of selection or sampling weights (IPSW) (1, 3-7). IPSW reweights study participants to resemble a chosen target population on observed covariates, providing an estimator whose output is an estimate of the population average treatment effect (PATE) (3-5, 7). IPSW is heavily reliant on three conditions: the careful selection of the target population, the availability of relevant data, and the inclusion of appropriate covariates in the weighting procedure (3-5, 7). If the target population is mis-specified or not truly representative of the individuals to whom the treatment effect should be generalized, then the marginal covariate distributions will differ meaningfully from those in the intended population. Under these conditions, IPSW produces an estimate of the PATE for the population represented in the available dataset rather than the true target population. In this sense, IPSW remains unbiased for the estimand defined by the available dataset, but that estimand may differ meaningfully from the PATE for the population of substantive interest. This disconnect can lead to apparent bias when results are interpreted with respect to the intended target population, which can misinform downstream decision-making. While weighting has the potential to increase the generalizability of treatment effect estimates, the risk of biased results is worth exploring in detail to see the impact of overlooking the choice of target population data in practice.

In this study, we design simulations based on real data to demonstrate how weighting to non-representative populations can yield PATE estimates that differ from those for the intended target population. We quantify the magnitude of differences, which will be referred to as bias, across scenarios with varying degrees of treatment effect modification. Our simulation draws on empirical data distributions from real-world sources—including trial data, disease registries, and national population summaries—that together span a conceptual continuum from highly selected to broadly inclusive populations. We begin by introducing our motivating example in type II diabetes, then outline the conceptual spectrum of potential target populations used for generalization and conclude with simulations evaluating how IPSW performance and PATE bias vary across these populations.

**Data Application: Type II Diabetes**

Although the framework we describe applies broadly across clinical and non-clinical settings, we ground our discussion in type II diabetes to provide a concrete illustration. We consider the goal of generalizing a trial-based estimate of semaglutide's treatment effect on percent weight change to broader populations of patients with type II diabetes (10). This example reflects a common scenario in comparative effectiveness research where inference is desired from a targeted analytic cohort to a broader group. The magnitude of the treatment effect used in our simulation was informed by this trial to anchor the simulations to realistic scenarios for research teams.

**Candidate Target Populations**

*Figure 1: Conceptual Spectrum of Target Populations*

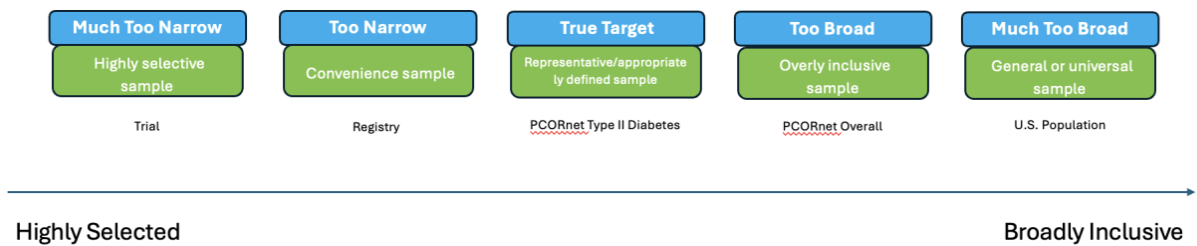

A conceptual spectrum of target populations helps to illustrate the importance of careful selection of a target population dataset for generalization (*Figure 1*). At one extreme, the target population may be defined too narrowly, including only highly selected samples and ignoring the need for generalizability. Moving outward, a target population may be somewhat broader but still restrictive, capturing only a subset of the true population while excluding others due to selection biases. At the center lies the "true" target population. This is the population that best approximates the intended inference population on relevant covariates. Extending further, a population that is too broad includes both relevant and irrelevant individuals, diluting estimates of the true effect. Finally, at the broadest extreme, the population may be so inclusive that it encompasses nearly all individuals, most of whom are not candidates for the treatment.

In this study, we anchor this spectrum to five realistic datasets commonly available to applied researchers studying type II diabetes:

- **Highly selected analytic sample (much too narrow):** Based on a study sample, which includes only highly selected individuals meeting strict eligibility criteria. This is typically the study that we are starting with and hoping to generalize – in other words, the sample before weighting. In this simulation study, a randomized clinical trial (RCT) serves as a motivating and illustrative example, although the same principles apply to any analytic design where results are generalized beyond the study sample. To simulate our trial population, covariate distributions were modeled on the *Kidney and Cardiovascular Effectiveness of Empagliflozin Compared With Dipeptidyl Peptidase-4 Inhibitors in Patients With Type 2 Diabetes* study (9), and outcomes were based on *One-year Weight Reduction with Semaglutide or Liraglutide in Clinical Practice* (10).

- **Restrictive convenience sample (too narrow):** Based on disease-specific registries that capture clinical data from select health centers but may omit broader community populations. Our convenience sample data were based on the DCR, as this registry contains information on type II diabetics, though the majority of records originate from cardiology practices as of late 2015 (11).
- **Reference sample (true target):** Based on a multi-health-system dataset that best approximates the true population of patients with type II diabetes, providing a clinically relevant and geographically diverse source for generalization. In this study, the PCORnet type II diabetes population was used, as this represents a robust data source that captures medical information on type II diabetics across medical specialties, health systems, and geographic locations. The reference population was parameterized using internal summaries from PCORnet by Dr. Keith Marsolo and Dr. Darcy Louzao (12).
- **Overly inclusive sample (too broad):** Based on the total patient population in a multi-health-system dataset. In this study, the PCORnet overall population was used, as this population includes patients with a wide range of conditions, or lack thereof, thereby diluting our type II diabetes-specific reference population. The overly inclusive population was parameterized using internal summaries from PCORnet by Dr. Keith Marsolo and Dr. Darcy Louzao (12).
- **General population (much too broad):** Based on national census data, of which the vast majority of people do not have the disease of interest and are not candidates for treatment. In this study, the general population was based on the United States population, as this population includes nearly all individuals, the majority of which do not have type II diabetes or are eligible for treatment. This population was parameterized

> with only demographic parameters, which were drawn from the *U.S. Census Bureau's April 1, 2020 population estimates.*

Notably, these empirical data were used to anchor our simulation in realistic magnitudes of population differences, not to assert that any specific dataset represents truth.

This conceptual model underscores the key choice inherent in conducting IPSW: a target population dataset. In practice, analysts may have a clearly defined population for which inference is desired, but are constrained by the datasets available to operationalize that target. Any of the above candidates may therefore be deemed reasonable for the generalization of a study treatment effect estimate, depending on the context. However, each corresponds to a different representation of the target population and thus a different estimand. Consequently, these choices have substantially different implications – implications that we explore through comparing the results of IPSW fit to each candidate data set.

**IPSW**

For reference, the sample average treatment effect (SATE) for the realized study sample is defined as:

$$SATE = \frac{1}{n_s} \sum_{i:S_i=1} (Y_i(1) - Y_i(0))$$

where $n_s$ is the number of trial participants.

Additionally, the population-level PATE is defined as:

$$PATE = \frac{1}{N} \sum (Y_i(1) - Y_i(0))$$

To generalize study estimates, we focus in on the technique of IPSW. Let $S_i = 1$ indicate trial participation and $S_i = 0$ indicate membership in the target population, with $X_i$ denoting the vector of covariates (1, 3). The probability of trial inclusion, $p(S = 1 | X_i)$, can be estimated using a logistic regression model. From this model, each trial participant receives a sampling weight:

$$w_i = \frac{p(S = 1)}{p(S = 1 | X_i)}$$

The PATE is then estimated as the weighted average of individual treatment effects:

$$PATE = \frac{\Sigma(w_i \, TE_i)}{\Sigma(w_i)} = \frac{\Sigma(w_i \, [Y_i(1) - Y_i(0)])}{\Sigma(w_i)}$$

where $TE_i = Y_i(1) - Y_i(0)$.

In this framework, IPSW estimates the PATE for a specified target population by reweighting the trial to match that target on observed covariates.

**Simulation Study**

We conducted a Monte Carlo simulation study to evaluate the performance of IPSW estimators of the PATE when generalizing randomized trial findings to a variety of candidate target populations. Each iteration simulated a highly selected analytic population (9, 10) and four distinct target populations—convenience sample (11), reference (12), overly inclusive (12), and general (13)—representing a continuum from narrowly defined analytic samples to broadly inclusive national populations. Population sizes were fixed at 5,000 for the analytic sample, 75,000 for the convenience sample, 150,000 for the reference sample, 300,000 for the overly inclusive sample, and 500,000 for the general population.

Data sources defined the distributions for age, sex, race, Hispanic ethnicity, hypertension, heart failure, coronary artery disease (CAD), and peripheral artery disease (PAD). Of note, the general population (US Census) only contained demographic characteristics. Covariates were selected based on variables that were consistently available and comparable across data sources, ensuring representation in all population levels. We grouped these variables into demographic (age, sex, race, Hispanic ethnicity), and clinical (hypertension, heart failure, coronary artery disease, peripheral artery disease) domains to reflect their distinct roles in population characterization and treatment effect modification.

To clarify how empirical data informed our design, we directly controlled all simulated populations so that they aligned along a pre-specified continuum of representativeness—from highly selected to broadly inclusive groups. The ordering of populations along this continuum was based on knowledge of how each data source was collected and defined, moving from the most highly selective datasets, such as clinical trials and disease registries, to the most broadly inclusive sources, such as national population data. The PCORnet disease-area population served as our reference cohort because it provided the most robust and comprehensive representation of the intended inference population available to us. However, we make no assumption that this population reflected the true marginal distribution of patients with type II diabetes in the broader United States or any empirical "truth." Example datasets were assigned to play the role of "too selective" or "too broad" based on knowledge that their inclusion and exclusion criteria were more restrictive or more inclusive than those of the reference population. Real-world data sources were therefore used solely to motivate the magnitude of population differences that would be realistic in applied research, while the continuum of representativeness itself was fully controlled within the simulation.

Before applying IPSW to obtain estimates of the PATE, covariate differences between populations were quantified using absolute standardized mean differences (SMDs) calculated relative to the reference population. These were visualized using Love plots, providing a graphical representation of population-level imbalance.

Potential outcomes were generated from linear models representing untreated and treated potential outcomes for each simulated individual. The untreated potential outcome, $Y_i(0)$, followed:

$$Y_i(0) = \mu_0 + \beta^T X_i + \varepsilon_i$$

where $\mu_0 = 3.1$ represented the expected control outcome in patients with type 2 diabetes based on *Gasoyan et al.* (10), and all covariate effects ($\beta_j$) were fixed at –0.50. Random error $\varepsilon_i$ was drawn from a normal distribution with mean 0 and standard deviation 7, which was consistent with the observed variability in weight change outcomes in the same trial.

The treated potential outcome, $Y_i(1)$, was defined as:

$$Y_i(1) = \mu_1 + (\beta + \delta)^T X_i + \varepsilon_i$$

where $\mu_1 = \mu_0 + 5.4$, corresponding to a Cohen's d of 0.8 and representing a large treatment effect scaled from the effect of semaglutide relative to liraglutide reported in *Gasoyan et al.* (10). Cohen's d is a standardized measure of mean difference, defined as the difference in group means divided by the pooled standard deviation, and is commonly expressed as small ($\approx 0.2$), medium ($\approx 0.5$) and large ($\approx 0.8$) following Cohen's conventions (8). Treatment-effect modifier coefficients ($\delta_j$) were all set to 1.34, corresponding to a *Cohen's d* of 0.2, to induce small treatment-effect shifts associated with each modifying covariate. Treatment-related parameters of treatment effect and covariate-specific moderation effects were defined based on Cohen's d, using pooled standard deviation of type II diabetic outcomes from *Gasoyan et. al* (10). In this setup, individuals possessing a given characteristic (e.g., having PAD, hypertension, or being female) experienced a modest increase in treatment effect relative to those without that characteristic, thereby generating controlled heterogeneity in the treatment response.

Four heterogeneity structures were evaluated to assess the impact of varying degrees of effect modification:

1) All modifiers: All demographic and clinical variables – age, sex, race, Hispanic ethnicity, hypertension, heart failure, peripheral artery disease (PAD), and coronary artery disease (CAD) – modified the treatment effect
2) Four modifiers: Only age, sex, hypertension, and PAD acted as treatment effect modifiers
3) Single Modifier: Only hypertension status modified treatment response
4) No modifiers: No variables modified treatment response (i.e., treatment effect was constant across all individuals).

IPSW was applied to obtain estimates of the PATE for each target population. Two weighting specifications were compared: a Demographic + Clinical model including all covariates (age, sex, race, Hispanic ethnicity, hypertension, heart failure, CAD, PAD), and a Demographic-only model including age, sex, race, and Hispanic ethnicity. Bias for each target population was calculated as the difference between the IPSW PATE estimate for that population and the reference PATE corresponding to the PCORnet disease-specific cohort under the Demographic + Clinical specification. Each simulation was repeated 1,000 times. To assess sensitivity to effect size, secondary analyses scaled the treatment effect and covariate shift from 0.5X to 2.5X, maintaining all other model parameters. All simulations and analyses were conducted in R version 4.4.0 using the tidyverse, gt, and base R functions. Outputs included mean and standard deviation summaries for Y(0), Y(1), SATE, IPSW-based PATE estimates, bias distributions, and visual comparisons of bias across target populations.

**Results**

*Table 1: Clinical and Demographic Characteristics of Target Populations*

**Table 1. Baseline Characteristics Across Populations**

Age shown as mean (SD). Categorical variables shown as count (percent).

|  | Trial | DCR | PCORnet disease-area | PCORnet overall | US Census |
|---|---|---|---|---|---|
| N | 62,197 | 902,772 | 4,233,764 | 41,619,147 | 331,515,736 |
| N (simulated) | 5,000 | 75,000 | 150,000 | 300,000 | 500,000 |
| Age (years) | 61.8 (12.7) | 67.9 (12.6) | 63.0 (13.7) | 41.4 (22.2) | 39.1 (23.5) |
| Sex (female) | 2518 (50.4%) | 34143 (45.5%) | 78082 (52.1%) | 170677 (56.9%) | 254255 (50.9%) |
| Race: White | 3554 (71.1%) | 63984 (85.3%) | 98684 (65.8%) | 209817 (69.9%) | 379679 (75.9%) |
| Race: Black | 1123 (22.5%) | 8767 (11.7%) | 34732 (23.2%) | 50012 (16.7%) | 67729 (13.5%) |
| Race: Other | 323 (6.5%) | 2249 (3.0%) | 16584 (11.1%) | 40171 (13.4%) | 52592 (10.5%) |
| Hispanic Ethnicity | 452 (9.0%) | 4075 (5.4%) | 23943 (16.0%) | 44948 (15.0%) | 93308 (18.7%) |
| Hypertension | 3919 (78.4%) | 65388 (87.2%) | 115463 (77.0%) | 71769 (23.9%) | — |
| Heart failure | 523 (10.5%) | 20134 (26.8%) | 22422 (14.9%) | 8879 (3.0%) | — |
| Coronary artery disease | 1150 (23.0%) | 43460 (57.9%) | 37310 (24.9%) | 21131 (7.0%) | — |
| Peripheral artery disease | 462 (9.2%) | 11594 (15.5%) | 34393 (22.9%) | 14918 (5.0%) | — |

*Figure 2: Standardized Mean Differences of Covariates Relative to PCORnet Disease-Area Population*

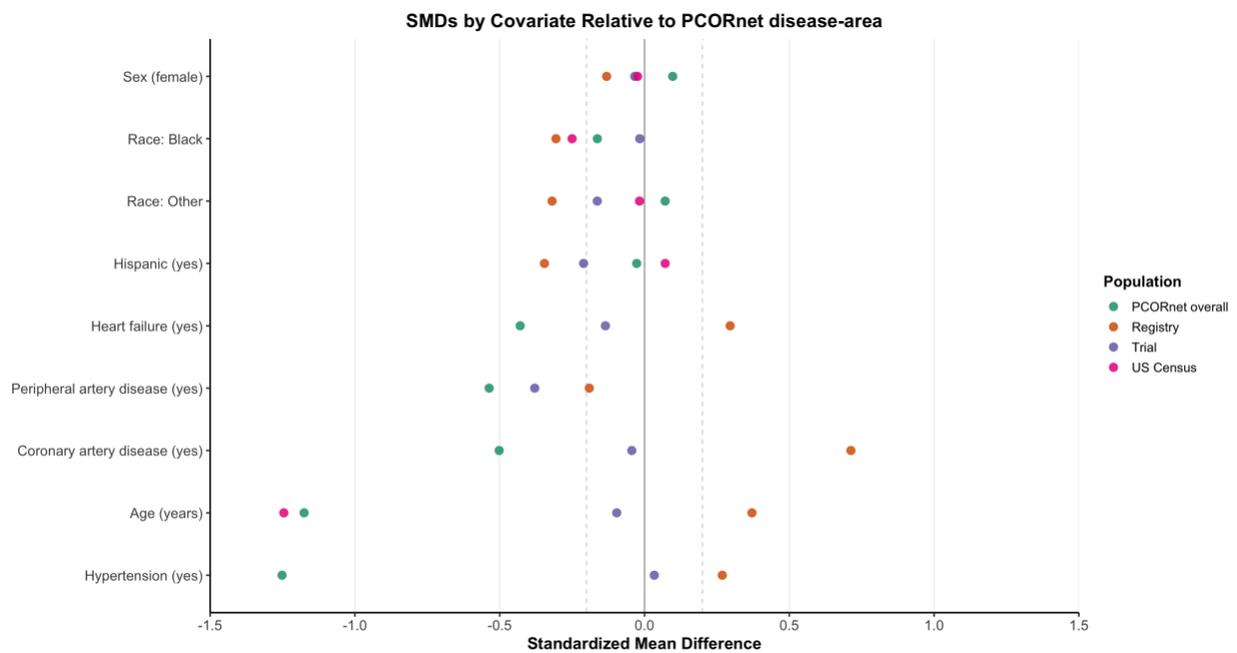

Baseline demographic and clinical characteristics for the highly selected analytic (trial), convenience sample (registry), reference (PCORnet disease-area), overly inclusive (PCORnet overall), and general (US census) populations are summarized in *Table 1*, with standardized mean differences relative to our reference population displayed in *Figure 2*. In this figure, populations with points that lie to the left of the vertical line at zero represent younger, less comorbid, less diverse populations while those with points to the right represent older, more comorbid and more diverse populations. Together, these results illustrate the systematic demographic and clinical differences across populations.

As shown in *Figure 2*, age exhibited one of the most pronounced differences in distribution across populations, with overly inclusive and general populations differing most notably from the reference population. Corresponding values in *Table 1* show that the mean age in the reference population (63.0 years) was similar to that in the highly selected analytic (61.8 years) population, and slightly younger than the convenience sample (67.9 years) population. However, the reference population was notably older than the broadly inclusive (41.4 years) and general (39.1 years) populations.

Additionally, *Figure 2* exhibits notable differences in the distribution of hypertension status across populations with the convenience sample displaying the greatest rightward deviation from the reference and the overly inclusive population displaying the greatest leftward deviance. Corresponding values in *Table 1* show that hypertension prevalence in the convenience sample was 87.2% compared with 77% in our reference population and 78.4% in the highly selected analytic population. Alternatively, the overly inclusive population had a hypertension prevalence of 23.9%, representing a much less comorbid population than our reference.

Finally, *Figure 2* shows that CAD prevalence had pronounced differences across populations with the convenience sample displaying the greatest rightward deviation and the overly inclusive population having the greatest leftward deviation from the reference. As reported in *Table 1*, the prevalence of CAD in the convenience sample was 57.9%, which is much greater than in the reference population (24.9%). Furthermore, the highly selected analytic sample (23%) was very similar to the reference population, while the overly inclusive population (7%) had a much lower prevalence.

*Figure 3: Bias Distributions of SATE and PATE Estimates Across Target Populations (All Covariates as Modifiers, 1X TE)*

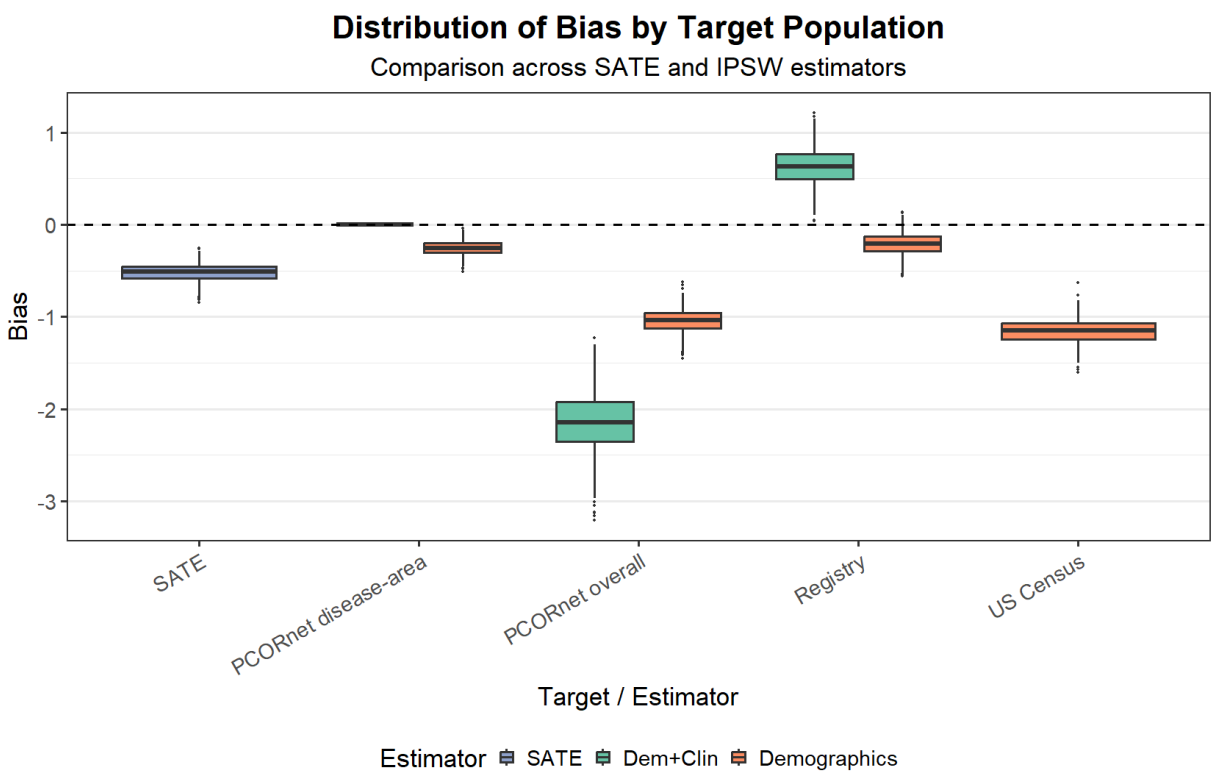

*Target populations span a conceptual spectrum of representativeness from a highly selected analytic sample (trial), convenience sample (registry), reference (PCORnet disease-area), overly inclusive (PCORnet overall), and general (U.S. Census) population. Two weighting models were used: Demographic + Clinical (all covariates) and Demographic-only (age, sex, race, ethnicity).*

*Figure 3* shows the distribution of bias in SATE and IPSW PATE estimates across target populations when all covariates modified the treatment effect. The reference population estimator using the Demographic + Clinical specification exhibited no bias, as this was our comparison estimate.

In contrast, weighting to less representative target populations using demographic and clinical variables introduced systematic and directionally consistent bias. Estimates targeting the convenience sample population were biased upward, reflecting generalization to an older, sicker cohort with higher expected treatment responses. Estimates targeting the overly inclusive population were biased downward, reflecting generalization to a younger, healthier group with lower expected effects. Across nonreference populations, bias magnitude increased as representativeness decreased, mirroring the imbalance patterns observed in *Figure 1*. This shows that even under correct model specification, weighting to nonrepresentative targets produced bias in PATE estimates, underscoring that the validity of IPSW generalization depends heavily on target population identification.

Additionally, the SATE estimate was biased downward, consistent with expectations for a younger, healthier, and less comorbid highly selected analytic sample population relative to the reference cohort. Notably, the magnitude of SATE bias was smaller than that observed for several nonrepresentative target populations, illustrating that weighting to an inappropriate target can introduce more bias than omitting weighting altogether. Furthermore, bias patterns differed by weighting specification. Demographic-only weighting generally produced smaller absolute bias than the Demographic + Clinical model. This attenuated bias occurred because demographic variables captured only part of the true effect modification structure. The same trend as noted above is observed in the demographic-only weighting model, as target populations that are increasingly nonrepresentative produced greater bias in PATE estimates.

*Figure 4: Bias Distributions of SATE and PATE Estimates Across Target Populations (Four Covariates as Modifiers, 1X TE)*

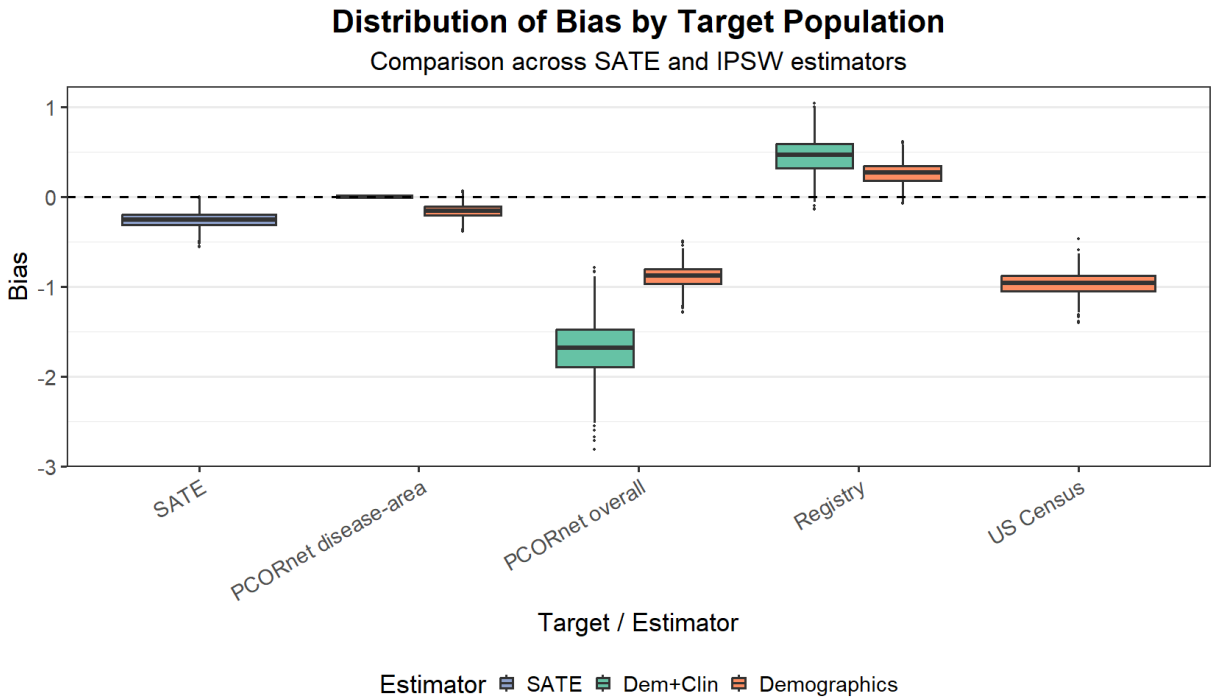

*Target populations span a conceptual spectrum of representativeness from a highly selected analytic sample (trial), convenience sample (registry), reference (PCORnet disease-area), overly inclusive (PCORnet overall), and general (U.S. Census) population. Two weighting models were used: Demographic + Clinical (all covariates) and Demographic-only (age, sex, race, ethnicity).*

*Figure 4* displays the bias distributions for SATE and PATE estimates across target populations when only age, sex, hypertension, and peripheral artery disease modified the treatment effect. The reference population (Demographic + Clinical) estimate again exhibited negligible bias. As in the prior scenario, bias magnitude increased with decreasing representativeness of the target population. Estimates for the convenience sample remained biased upward, reflecting generalization to an older, more comorbid cohort with higher expected treatment responses, while estimates for the overly inclusive population were biased downward, reflecting generalization to a younger, healthier population. The SATE estimate was again biased downward, consistent with a younger and less comorbid trial cohort, but its bias remained smaller than that observed for several nonrepresentative PATE estimates, illustrating that weighting to an inappropriate target population can be more harmful than not weighting at all.

When weights were constructed using Demographic-only models, bias patterns persisted but were somewhat attenuated. The convenience sample and overly inclusive populations displayed smaller absolute bias compared to the Demographic + Clinical specification, though this reduction did not reflect improved performance. Rather, the demographic-only model failed to account for all true effect modifiers—particularly clinical covariates such as hypertension and peripheral artery disease—thereby producing estimates that appeared closer to zero only because part of the heterogeneity was ignored. The general population, which contained only demographic information, remained substantially biased, reflecting its poor correspondence to the disease-specific cohort.

*Figure 5: Bias Distributions of SATE and PATE Estimates Across Target Populations (One Covariate as Modifiers, 1X TE)*

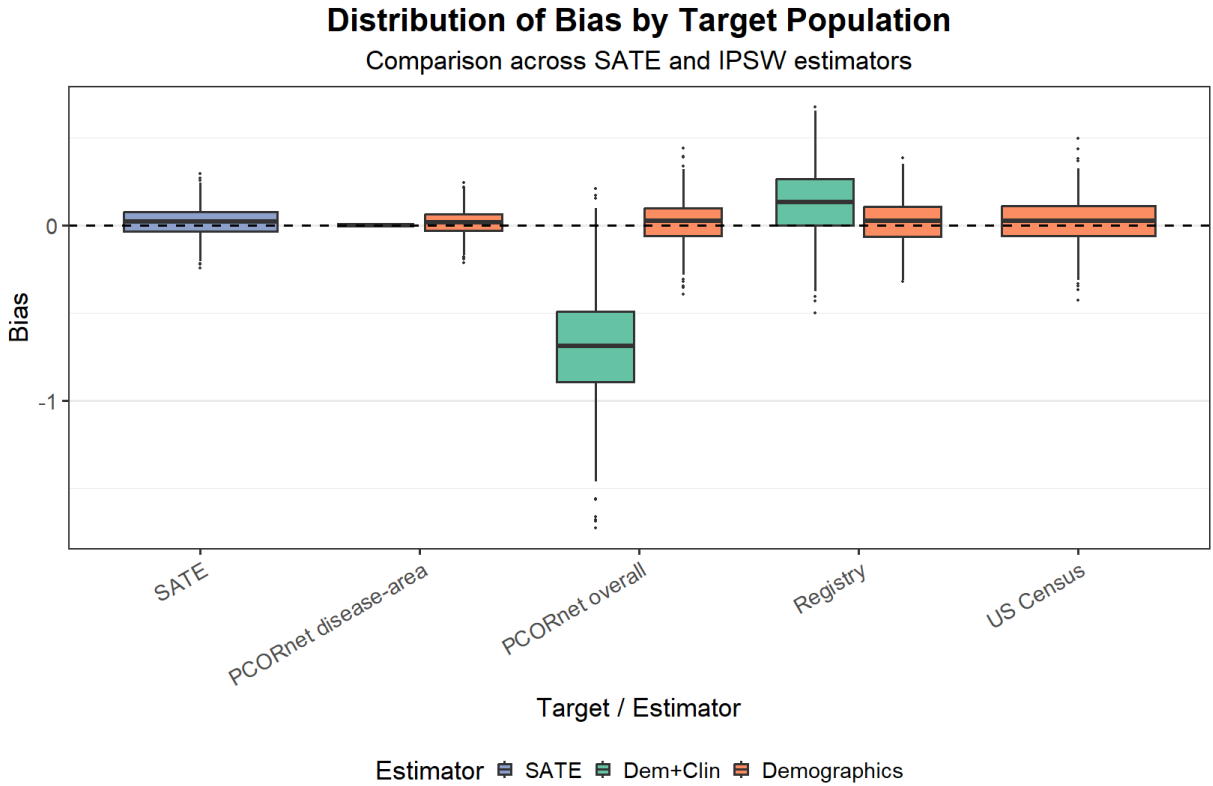

*Target populations span a conceptual spectrum of representativeness from a highly selected analytic sample (trial), convenience sample (registry), reference (PCORnet disease-area), overly inclusive (PCORnet overall), and general (U.S. Census) population. Two weighting models were used: Demographic + Clinical (all covariates) and Demographic-only (age, sex, race, ethnicity).*

*Figure 5* presents bias distributions across target populations when only hypertension acted as a treatment-effect modifier. This scenario isolates the influence of a single modifying covariate, allowing us to examine how differences in that variable's prevalence alone generate bias in IPSW PATE estimates. The reference population (Demographic + Clinical) estimate again showed negligible bias. Estimates for the convenience sample were biased slightly upward, consistent with its higher prevalence of hypertension and greater comorbidity burden, whereas estimates for the overly inclusive population were biased downward, reflecting generalization to a population with substantially lower hypertension prevalence. The SATE did not show pronounced bias, as the reference and highly selected analytic populations were very similar with respect to hypertension status. Again, we observe that the SATE estimate produces a lesser degree of bias than several other PATE estimates, reinforcing that weighting to a misaligned target can be more harmful than not weighting at all.

When weights were constructed using the Demographic-only specification, bias approximately equivalent to that of the SATE was observed across populations, as none of the included demographic covariates were true effect modifiers. This scenario highlights that when the treatment effect is modified by a single non-demographic variable, excluding that variable from the weighting model yields uniformly unbiased estimates, but for the wrong reason. The model simply fails to capture any heterogeneity. Consequently, the absence of bias does not indicate correctness, but rather reflects that the weighting model is insensitive to the true source of variation. Together, these results illustrate that bias magnitude is directly proportional to the extent of population imbalance in the modifying variable, and emphasize the need for inclusion of clinically relevant variables in weighting models.

*Figure 6: Bias Distributions of SATE and PATE Estimates Across Target Populations (No Covariates as Modifiers, 1X TE)*

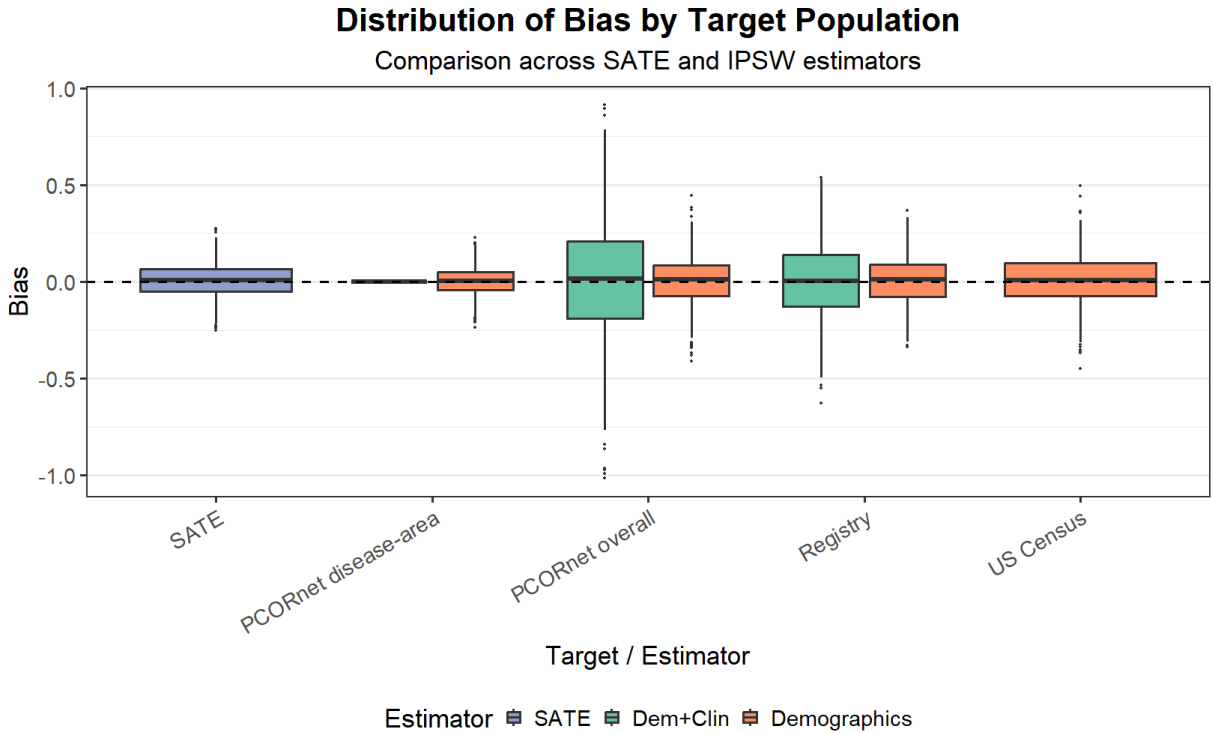

*Target populations span a conceptual spectrum of representativeness from a highly selected analytic sample (trial), convenience sample (registry), reference (PCORnet disease-area), overly inclusive (PCORnet overall), and general (U.S. Census) population. Two weighting models were used: Demographic + Clinical (all covariates) and Demographic-only (age, sex, race, ethnicity).*

*Figure 6* presents bias distributions when no covariates modified the treatment effect, representing a constant treatment effect across all individuals. As expected, all estimators converged to the true PATE, and no systematic bias was observed across populations or weighting specifications. The SATE, Demographic + Clinical, and Demographic-only estimators each yielded unbiased results, with bias distributions centered near zero and similar variability. This scenario confirms that when treatment effects are truly homogeneous, IPSW performs equivalently regardless of model specification or target population, as there is no heterogeneity for population differences to distort.

Importantly, this finding reinforces the results from prior scenarios and from existing theoretrical results: bias arises only when treatment effects vary across covariates whose distributions differ between trial and target populations. When no such heterogeneity exists, the choice of weighting variables or target population becomes inconsequential, and all estimators recover the same unbiased treatment effect.

**Discussion**

This study demonstrates that the accuracy of IPSW estimators of the PATE depends critically on target population representativeness. Across all simulation scenarios, greater differences in observed covariate distributions between a target population and the reference population were associated with larger bias in IPSW PATE estimates. Populations that were either overly restrictive or overly inclusive produced systematically distorted estimates, even when the IPSW procedure was correctly specified.

These findings highlight that IPSW performance is partially driven by how well the target population represents the population for which inference is intended. Weighting can adjust a highly restricted sample to resemble a target population, but it cannot recover the true PATE when that population differs meaningfully from the one underlying the desired inference. Directional bias followed predictable patterns, including overestimation when generalizing to older or more comorbid populations, and underestimation when generalizing to younger or healthier ones.

The conceptual spectrum of target populations further illustrates this relationship. At one extreme, narrowly defined populations exclude relevant subgroups, limiting generalizability and leading to biased estimates that reflect only a subset of the intended population. At the other end, overly inclusive populations dilute the treatment effect by incorporating individuals who differ substantially from the context of interest.

This simulation study was designed to isolate the role of target population representativeness under controlled conditions. While empirical covariate distributions enhanced realism, several simplifying assumptions were necessary. Covariates were simulated independently, and model specification was idealized to ensure that observed bias reflected population differences. In applied settings, additional complexities such as correlated covariates, missing data, and unobserved effect modifiers may introduce further bias or variability. Moreover, this framework focused on a single disease area and set of covariates; the patterns and magnitudes of bias may differ across other clinical domains or data structures. Finally, we did not evaluate alternative generalization approaches (e.g., doubly robust or Bayesian estimators), which may perform differently under similar conditions.

**Conclusions**

Bias increased systematically as target populations diverged from a well-representative population, underscoring that the accuracy of IPSW-based generalization is determined partly by the representativeness of the chosen target. Statistical reweighting can align a restricted sample with a specified population but cannot correct for fundamental differences in who that population represents. Accordingly, defining and justifying the target population should be regarded as a central design decision in generalization studies.

**Appendix**

*Table 2: Signed Standardized Mean Differences by Covariate*

**Signed SMDs by Covariate and Population**

Reference = PCORnet disease-area (not shown).

| | Trial | DCR | PCORnet overall | US Census |
|---|---|---|---|---|
| Age (years) | −0.096 | 0.371 | −1.176 | −1.246 |
| Sex (female) | −0.034 | −0.131 | 0.097 | −0.024 |
| Race: Black | −0.017 | −0.306 | −0.163 | −0.250 |
| Race: Other | −0.163 | −0.319 | 0.071 | −0.017 |
| Hispanic Ethnicity | −0.210 | −0.346 | −0.027 | 0.071 |
| Hypertension | 0.034 | 0.269 | −1.252 | — |
| Heart failure | −0.135 | 0.296 | −0.429 | — |
| Coronary artery disease | −0.044 | 0.713 | −0.502 | — |
| Peripheral artery disease | −0.379 | −0.191 | −0.537 | — |
| ▶ All Covariates as Effect Modifiers — All Used | **−1.044** | **0.356** | **−3.917** | **−1.466** |
|     Demographic Only | — | *−0.731* | *−1.197* | *−1.466* |
| ▶ Age + Sex + Hypertension + PAD Effect Modifiers — All Used | **−0.475** | **0.318** | **−2.867** | **−1.270** |
|     Demographic Only (Age + Sex) | — | *0.240* | *−1.078* | *−1.270* |
| ▶ Hypertension Only Effect Modifier — All Used | **0.034** | **0.269** | **−1.252** | **0.000** |
| ▶ No Effect Modifiers (All Zero) | **0.000** | **0.000** | **0.000** | **0.000** |

*Table 3: Simulation results of SATE and PATE Estimates (All Covariates as Modifiers, 1X TE)*

| Weighting Model | Target | PATE (mean) | PATE (SD) | Bias (mean) | Bias (SD) |
|---|---|---|---|---|---|
| | SATE | 8.197 | 0.149 | −0.519 | 0.093 |
| Dem+Clin | PCORnet disease-area | 8.716 | 0.167 | 0.000 | 0.000 |
| Dem+Clin | Registry | 9.350 | 0.246 | 0.634 | 0.200 |
| Dem+Clin | PCORnet overall | 6.571 | 0.323 | −2.145 | 0.308 |
| Demographics | PCORnet disease-area | 8.465 | 0.155 | −0.251 | 0.076 |
| Demographics | Registry | 8.508 | 0.172 | −0.208 | 0.126 |
| Demographics | PCORnet overall | 7.676 | 0.180 | −1.041 | 0.128 |
| Demographics | US Census | 7.565 | 0.183 | −1.152 | 0.131 |

SATE & PATE (×1)
Bias = estimate − PCORnet disease-area PATE (Dem+Clin)

*Table 4: Simulation results of SATE and PATE Estimates (Four Covariates as Modifiers, 1X TE)*

| Weighting Model | Target | PATE (mean) | PATE (SD) | Bias (mean) | Bias (SD) |
|---|---|---|---|---|---|
| | SATE | 7.258 | 0.148 | −0.260 | 0.091 |
| Dem+Clin | PCORnet disease-area | 7.518 | 0.166 | 0.000 | 0.000 |
| Dem+Clin | Registry | 7.974 | 0.244 | 0.456 | 0.197 |
| Dem+Clin | PCORnet overall | 5.829 | 0.322 | −1.688 | 0.306 |
| Demographics | PCORnet disease-area | 7.359 | 0.155 | −0.159 | 0.075 |
| Demographics | Registry | 7.781 | 0.172 | 0.263 | 0.125 |
| Demographics | PCORnet overall | 6.633 | 0.180 | −0.885 | 0.126 |
| Demographics | US Census | 6.551 | 0.183 | −0.966 | 0.129 |

SATE & PATE (×1). Bias = estimate − PCORnet disease-area PATE (Dem+Clin)

*Table 5: Simulation results of SATE and PATE Estimates (One Covariate as Modifier, 1X TE)*

| Weighting Model | Target | PATE (mean) | PATE (SD) | Bias (mean) | Bias (SD) |
|---|---|---|---|---|---|
| | SATE | 6.454 | 0.148 | 0.022 | 0.087 |
| Dem+Clin | PCORnet disease-area | 6.433 | 0.165 | 0.000 | 0.000 |
| Dem+Clin | Registry | 6.567 | 0.239 | 0.134 | 0.194 |
| Dem+Clin | PCORnet overall | 5.745 | 0.319 | −0.688 | 0.303 |
| Demographics | PCORnet disease-area | 6.451 | 0.153 | 0.019 | 0.074 |
| Demographics | Registry | 6.456 | 0.170 | 0.023 | 0.123 |
| Demographics | PCORnet overall | 6.457 | 0.179 | 0.024 | 0.123 |
| Demographics | US Census | 6.457 | 0.182 | 0.024 | 0.127 |

SATE & PATE (×1). Bias = estimate − PCORnet disease-area PATE (Dem+Clin)

*Table 6: Simulation results of SATE and PATE Estimates (No Covariates as Modifiers, 1X TE)*

**SATE & PATE (×1)**

Bias = estimate − PCORnet disease-area PATE (Dem+Clin)

| Weighting Model | Target | PATE (mean) | PATE (SD) | Bias (mean) | Bias (SD) |
|---|---|---|---|---|---|
| | SATE | 5.406 | 0.147 | 0.006 | 0.086 |
| Dem+Clin | PCORnet disease-area | 5.400 | 0.165 | 0.000 | 0.000 |
| Dem+Clin | Registry | 5.407 | 0.239 | 0.007 | 0.193 |
| Dem+Clin | PCORnet overall | 5.414 | 0.319 | 0.013 | 0.303 |
| Demographics | PCORnet disease-area | 5.403 | 0.153 | 0.003 | 0.073 |
| Demographics | Registry | 5.408 | 0.169 | 0.008 | 0.123 |
| Demographics | PCORnet overall | 5.409 | 0.178 | 0.009 | 0.123 |
| Demographics | US Census | 5.409 | 0.182 | 0.009 | 0.126 |